\documentclass[preprint]{revtex4-1}
\usepackage{graphicx}
\usepackage{dcolumn}
\usepackage{bm,bbm}
\begin{document}

\title{Optical finite representation of the Lorentz group}

\author{B. M. Rodr\'{\i}guez-Lara}
\affiliation{Instituto Nacional de Astrof\'{\i}sica, \'Optica y Electr\'onica, Calle Luis Enrique Erro No. 1, Sta. Ma. Tonantzintla, Pue. CP 72840, M\'exico}
\email{bmlara@inaoep.mx}

\author{J. Guerrero}
\affiliation{Departamento de Matem\'atica Aplicada, Facultad de Inform\'atica, Campus Espinardo, Univesidad de Murcia, 30100 Murcia, Spain.}

\begin{abstract}
We present a class of photonic lattices with an underlying symmetry given by a finite-dimensional representation of the 2+1D Lorentz group.
In order to construct such a finite-dimensional representation of a non-compact group, we have to design a $\mathcal{PT}$-symmetric optical structure.
Thus, the array of coupled waveguides may keep or break $\mathcal{PT}$-symmetry, leading to a device that behaves like an oscillator or directional amplifier, respectively.
We show that the so-called linear $\mathcal{PT}$-symmetric dimer belongs to this class of photonic lattices.
\end{abstract}

\maketitle

\section{Introduction}

Photonic lattices provide a solid platform for the simulation of relativistic and quantum physics \cite{Christodoulides2003p817,Longhi2009p243,Longhi2011p453,RodriguezLara2015p068014}.
The use of finite or pseudo-infinite arrays, with an adequate set of effective refractive indices and coupling parameters, allows the optical realization of compact, $SU(N)$, and non-compact groups, $SU(1,1)$, for example \cite{RodriguezLara2014p013802,RodriguezLara2014p2083,VillanuevaVergara2015p}.
The ability to classify photonic lattices by their underlying symmetry opens the door to the simulation of a large class of quantum phenomena and the use of symmetries to design integrated photonic devices.

Here, we are interested in the Lorentz group because it is ubiquitous in all linear theories where a state is represented as a wave function that belongs to a linear manifold with a scalar product, e.g., relativistic quantum mechanics, quantum mechanics, and electromagnetic theory in linear media \cite{Majorana1932p335,Dirac1936p447,Wigner1939p149,Dirac1945p284,Bargmann1947p568,Gelfand1968p1}. 
In optics, some matrix representations of the homogeneous Lorentz group have served as a bridge between the Jones and Mueller matrix calculus of polarization optics \cite{Takenaka1973p37,Simon1982p293,Han1997p2290,Kim2000p1,Franssens2015p164}, and have also been used in paraxial ray optics to describe  multi-lens sytems, interferometers, laser cavities and multilayer systems \cite{Baskal2004p455}.

The unitary and irreducible representations of the Lorentz group are well known \cite{Wigner1939p149,Bargmann1947p568,Gelfand1968p1}.
The Lorentz group is non-compact, thus, its unitary representations are infinite dimensional and we need infinite arrays of coupled lattices to create photonic lattices with this symmetry. 
In the laboratory, this means manufacturing an array with a large number of coupled waveguides and making sure propagated light never reaches one of the ends. 
The Lorentz group also has a family of finite-dimensional representations which are, however, non-unitary due to its non-compactness \cite{Bargmann1947p568,Barut1965p532}.
The best way to construct these finite representations is to start from the $su(2)$ Lie algebra with generators $\hat{J}_{j}$ and complexify it by defining $\hat{P}_{j} \equiv i \hat{J}_{j}$ with $j=0,x,y$.
They all close a new Lie algebra that is isomorphic to the Lorentz algebra in 3+1D, $so(3,1)$, containing various conjugated copies of the $so(2,1)$ Lie algebra, $\{\hat{K}_{0}, \hat{K}_{x}, \hat{K}_{y} \} \equiv \{\hat{J}_{0}, \hat{P}_{x}, \hat{P}_{y}\}, ~ \{ \hat{J}_{x}, \hat{P}_{y}, \hat{P}_{0}\}, ~\{\hat{J}_{y}, \hat{P}_{0}, \hat{P}_{x}\}$, that is, the Lorentz algebra in 2+1D.

Taking into account that the propagation of electromagnetic fields through a photonic lattice with underlying $SU(2)$ symmetry is modeled by the mode coupling matrix \cite{VillanuevaVergara2015p},
\begin{eqnarray}
\hat{H}_{SU(2)}(z) &=&  \omega \hat{J}_{0} + \lambda ( \hat{J}_{-} + \hat{J}_{+} ), \\
&=& \omega \hat{J}_{0} + 2 \lambda \hat{J}_{x},
\end{eqnarray}
we find $\{ \hat{J}_{x}, \hat{P}_{y}, \hat{P}_{0}\}$ as the natural choice for an optical finite dimensional representation of the 2+1D Lorentz group because it implies the change $\omega \rightarrow i \gamma$ in the effective refractive index parameter.
This translates into waveguides with identical real part of the effective refractive index and non-identical imaginary part, that is, linear gain or loss at each waveguide \cite{Somekh1973p46,Agarwal2012p031802}.  
In the following, we will study this optical finite representation of the Lorentz group.
First, we are going to show that it belongs to the class of  so-called $\mathcal{PT}$-symmetric optical systems.
Then, we will provide a closed-form propagator for the device in terms of a response to impulse function.
We will also show that the device behaves like an oscillator or directional amplifier for lattice parameters that keep or break the $\mathcal{PT}$-symmetry, in that order, and, finally, provide a working example in the form of the so-called linear $\mathcal{PT}$-symmetric dimer.

\section{Photonic lattice simulation}

Let us start with the coupled-mode differential set,
\begin{eqnarray}
-i \partial_{z} \vert \mathcal{E}(z) \rangle = \hat{H} \vert \mathcal{E} \rangle,
\end{eqnarray}
modeling the propagation of an electromagnetic field through an array of coupled waveguides. 
We have borrowed Dirac notation, such that column vectors are written as kets,
\begin{eqnarray}
\vert \mathcal{E}(z) \rangle &=& \left( \mathcal{E}_{0}(z), \mathcal{E}_{1}(z),  \ldots , \mathcal{E}_{2j}(z) \right)^{T}, \\
 &\equiv& \sum_{l=0}^{2j} \mathcal{E}_{l}(z) \vert j \rangle,
\end{eqnarray} 
where the operation $v^{T}$ stands for transposition, and mode coupling matrices as operators \cite{RodriguezLara2015p068014,VillanuevaVergara2015p}.
In particular, we are interested in waveguides with identical real part of the effective refractive index and linear gain or losses, that is, nonzero imaginary part of the effective refractive index, leading to the mode coupling matrix \cite{Somekh1973p46,Agarwal2012p031802,RodriguezLara2015p068014},
\begin{eqnarray}
\hat{H} &=& i \gamma \left( \hat{n} - j \right) + \lambda \left[ \hat{V} \sqrt{ \hat{n} (2j+1 - \hat{n}) } ~ + \right. \nonumber \\
&& \left. +  \sqrt{ \hat{n} (2j+1 - \hat{n}) } \hat{V}^{\dagger}  \right],\\
&\equiv& i \gamma \hat{J}_{0} + \lambda ( \hat{J}_{-} + \hat{J}_{+} ),
\end{eqnarray}
where the imaginary part of the effective refractive index and the effective coupling are given by the real parameters $\gamma$ and $\lambda$, in that order, and we have used the $SU(2)$ algebra that satisfies the commutation relations, $[\hat{J}_{+}, \hat{J}_{-}]=2 \hat{J}_{0}$ and $[\hat{J}_{0}, \hat{J}_{\pm}]= \pm \hat{J}_{\pm}$.
Also, we have used the step, $\hat{n} \vert j \rangle = j \vert j \rangle$, up-step, $\hat{V}^{\dagger} \vert j \rangle = \vert j+1 \rangle$, and down-step, $\hat{V} \vert j \rangle = \vert j-1 \rangle$ operators, that are diagonal, upper- and lower-diagonal square matrices of dimension $2j$, in that order.

First, we want to emphasize that the effective refractive index of the $n$-th and $(2j-n)$-th waveguides are complex conjugates of each other, with $j=0,1/2,1,3/2,\ldots$ and $2j+1$ is the number of waveguides in the device.
This is a signature of the so-called $\mathcal{PT}$-symmetric optical systems \cite{ElGanainy2007p2632,Guo2009p093902,Longhi2009p123601,Ruter2010p192,ElGanainy2013p161105,Longhi2014p1697}.
Thus, this photonic lattice has a dispersion relation,
\begin{eqnarray}
\Omega(m) = \Omega_{0}~(m-j), \quad \Omega_{0}=\sqrt{4 \lambda^2 - \gamma^{2}},
\end{eqnarray}
that may be real, $\gamma < 2 \lambda$, complete degenerate, $\gamma = 2 \lambda$, or purely imaginary, $\gamma > 2 \lambda$.

The impulse function of this array, that is, the field at the $n$-th waveguide given that the initial field impinged only at the $m$-th waveguide, can be constructed following the so-called Gilmore-Perelomov, symmetry based approach \cite{VillanuevaVergara2015p},
\begin{eqnarray}
\mathcal{I}_{m,n}(z)&=&  \langle n \vert e^{ \left[ - \gamma \hat{J}_{0} + i \lambda \left( \hat{J}_{+} + \hat{J}_{-} \right)\right]z} \vert m \rangle, \\
&=& \sqrt{ \left( \begin{array}{c} 2j \\ m \end{array}\right) \left( \begin{array}{c} 2j \\ n \end{array}\right)}  \frac{\left(  2 i \lambda \sin \frac{\Omega_{0}}{2} z  \right)^{m+n}}{ \Omega_{0}^{2j} } \times \nonumber \\
&& \times \left(  \Omega_{0} \cos \frac{\Omega_{0}}{2} z +  \gamma \sin \frac{\Omega_{0}}{2} z \right)^{2j-m-n} \times \nonumber \\
&&  \times ~K_{m}\left(n;\frac{4 \lambda^2}{ \Omega_{0}^{2}} \sin^{2}\frac{\Omega_{0}}{2} z, 2j  \right) ,
\end{eqnarray}
where the notation $\left(\begin{array}{c} a \\ b \end{array} \right)$ and $K_{n}(x,p,N)$ stand for the binomial coefficient and Krawtchouk polynomials \cite{Askey1975}.
This impulse function is valid for any given parameter regime.

\begin{figure}[htbp]
\centering
\includegraphics[scale = 1]{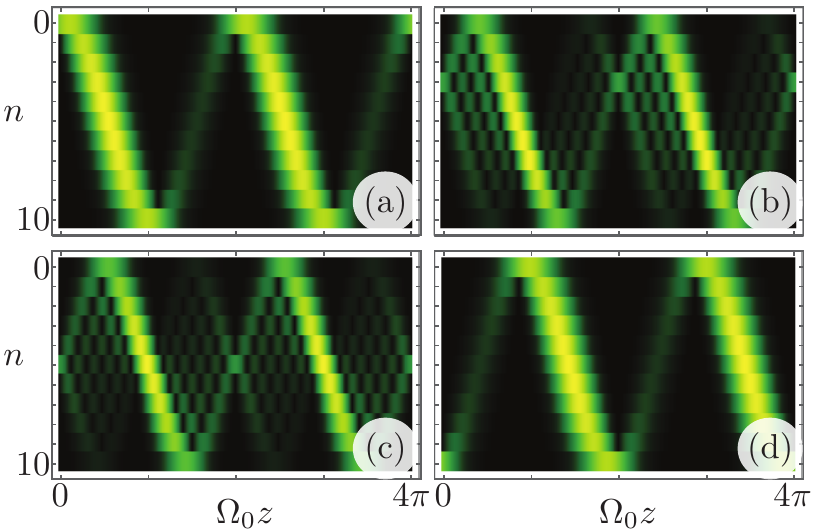}
\caption{(Color Online) Light intensity propagation, $\vert \mathcal{E}_{n}(z) \vert^2$, for an initial field amplitude impinging just the (a) $n=0$, (b)$n=3$, (c) $n=5$, (d) $n=10$ waveguides of a lattice in the $\mathcal{PT}$-symmetryc regime with parameters $\gamma = 0.4 \lambda$ and $j=5$.}\label{fig:Fig1}
\end{figure}

A device manufactured with parameters that keep the $\mathcal{PT}$-symmetry, $\gamma < 2 \lambda$, will show a $2 \pi / \Omega_{0}$ periodicity, Fig. \ref{fig:Fig1}. 
As we have said before, the fact that this is a non-unitary representation translates into a photonic lattice with gain and losses.
In practical terms, the $\mathcal{PT}$-symmetric optical system behaves like a directional amplifier or attenuator.
This is easily seen in Fig. \ref{fig:Fig1}(a) and Fig. \ref{fig:Fig1}(d), where an initial field impinging the zeroth, $m=0$, or last waveguide, $m=2j$, will produce a peak in the intensity at the symmetric waveguide, $n=2j-m$, at a propagation distance $z_{p} = (2k + 1) \pi / \Omega_{0}$ with $k=0,1,2,\ldots$,
\begin{eqnarray}
\mathcal{E}_{2j-m}(z_{p}) = \frac{1}{\Omega_{0}^{2}} \left(  2 i \lambda \sin \frac{\Omega_{0}}{2} z  \right)^{2j}, \quad m= 0, 2j.
\end{eqnarray}
In these particular cases it is simpler to see that for an input field impinging at the zeroth waveguide, $m=0$, the device behaves like an amplifier for half the propagation distance and then like an attenuator for the rest of the propagation.
Something similar occurs if the field impinges the last waveguide, $m=2j$, the device attenuates the input signal during the first half of the propagation length and then amplifies it during the second leg of the propagation.

\begin{figure}[htbp]
\centering
\includegraphics[scale = 1]{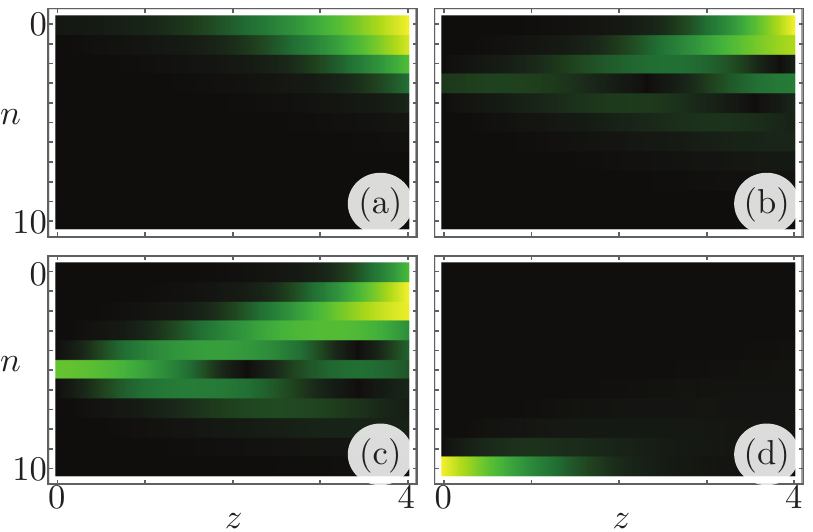}
\caption{(Color Online) Same as Fig \ref{fig:Fig1} for a completely degenerate lattice with parameters $\gamma = 2 \lambda$ and $j=5$.}\label{fig:Fig2}
\end{figure}

\begin{figure}[htbp]
\centering
\includegraphics[scale=1]{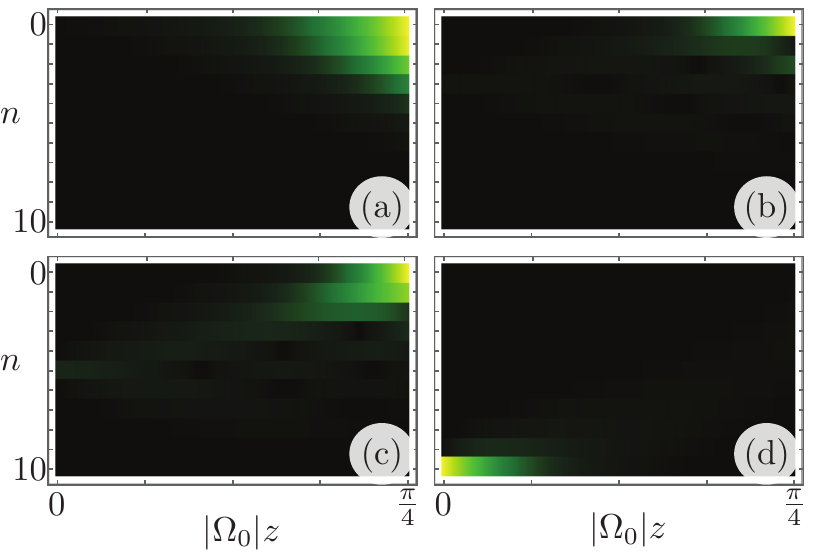}
\caption{(Color Online) Same as Fig \ref{fig:Fig1} for a completely degenerate lattice with parameters $\gamma = 2.5 \lambda$ and $j=5$.}\label{fig:Fig3}
\end{figure}

In the fully degenerate case, $\gamma = 2 \lambda$, it is straightforward to take the limit of the involved functions to find that the lattice looses its periodicity,
\begin{eqnarray}
\mathcal{I}_{m,n}(z)&=& \sqrt{ \left( \begin{array}{c} 2j \\ m \end{array}\right) \left( \begin{array}{c} 2j \\ n \end{array}\right)} \left( i \lambda z  \right)^{m+n} \left(  1 + \lambda z \right)^{2j-m-n} \times \nonumber \\
&&  \times ~K_{m}\left(n; \lambda^{2} z^{2}, 2j  \right).
\end{eqnarray}
Here, the device will work as a directional coupler,
concentrating most of the field amplitude from the zeroth to $j$-th lattices if enough propagation length is given, Fig. \ref{fig:Fig2}.
In the broken $\mathcal{PT}$-symmetry region, $\gamma > 2 \lambda$, the directional coupling with amplification just becomes more pronounced, Fig. \ref{fig:Fig3}.
Note that in all figures the brightest value corresponds to the maximum intensity in the array, which is not limited to the unit value due to the presence of amplification.
All figures were calculated via both a numeric solution of the mode coupling differential set and our analytic impulse function to good agreement.

\section{$\mathcal{PT}$-symmetric lineal dimer}

Now, we want to work out a practical example and consider the case with $j=1/2$,
\begin{eqnarray}
-i \partial_{z} \mathcal{E}_{0}(z)  &=&  -\frac{i}{2} \gamma \mathcal{E}_{0}(z)  + \lambda  \mathcal{E}_{1}(z), \\
-i \partial_{z} \mathcal{E}_{1}(z)  &=&  \frac{i}{2} \gamma \mathcal{E}_{1}(z)  + \lambda  \mathcal{E}_{0}(z).
\end{eqnarray}
These equations describe the linear $\mathcal{PT}$-symmetric dimer \cite{Guo2009p093902,Ruter2010p192}, therefore, it possesses an underlying $so(2,1)$ symmetry realized non-unitarily.
Note that we can uncouple the dynamics via a second derivative with respect to the propagation distance,
\begin{eqnarray}
\partial_{z}^2 \mathcal{E}_{j}(z) + \frac{1}{4} \Omega_{0}^2 \mathcal{E}_{j}(z) = 0, \quad j=0,1,
\end{eqnarray}
and discover three different behaviors. 
In the $\mathcal{PT}$-symmetric region, $\lambda > \gamma/2$, each field amplitude behaves like an oscillator with frequency $\Omega_{0}/2$, then, for lattice parameters $\lambda = \gamma / 2$, they exhibit free particle-like behavior and, finally, the field amplitudes behave like a repulsive oscillator in the broken symmetry regime, $\lambda < \gamma /2$.
The response to initial impulses are given by Eq.(9) and can be simplified in the following form,
\begin{eqnarray}
\mathcal{I}_{0,0}(z) &=& \cos \frac{\Omega_{0}}{2}z + \frac{\gamma}{\Omega_{0}} \sin  \frac{\Omega_{0}}{2}z, \\
\mathcal{I}_{0,1}(z) &=& \mathcal{I}_{1,0}(z)  , \\
&=&  \frac{i 2 \lambda}{\Omega_{0}}  \sin  \frac{\Omega_{0}}{2}z, \\
\mathcal{I}_{1,1}(z) &=& \cos \frac{\Omega_{0}}{2}z - \frac{\gamma}{\Omega_{0}} \sin  \frac{\Omega_{0}}{2}z. 
\end{eqnarray}

\begin{figure}[htbp]
	\centering
	\includegraphics[scale=1]{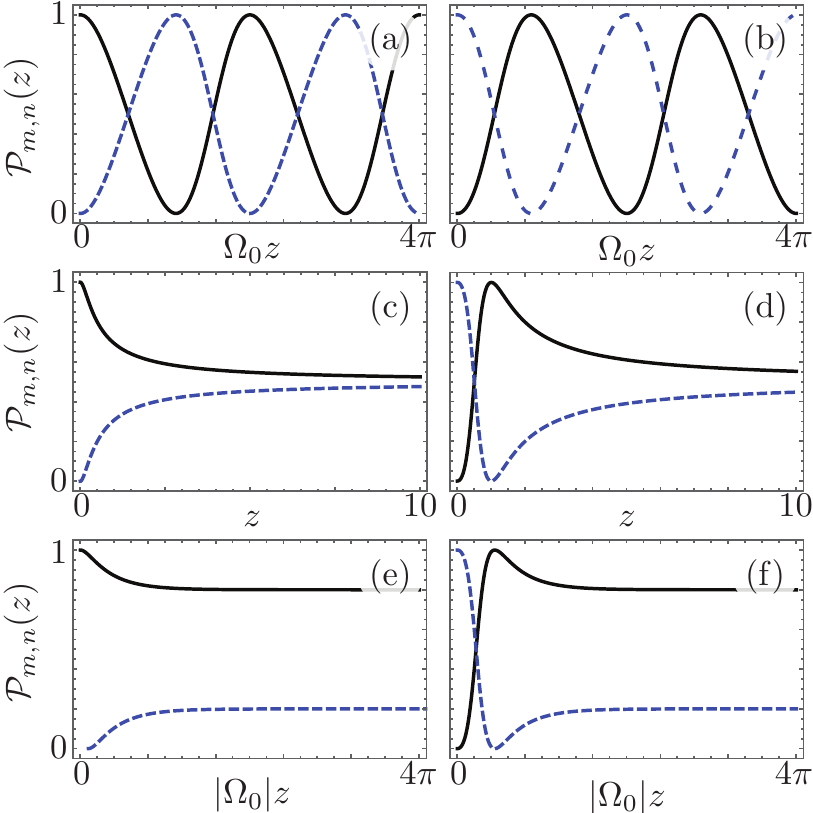}
	\caption{(Color Online) Re-normalized intensity, $\mathcal{P}_{m,n}(z)$, for a linear $\mathcal{PT}$-symmetric dimer with (a)-(b) $\gamma = 0.4 \lambda$, (c)-(d) $\gamma = 2 \lambda$ and (e)-(f) $\gamma = 2.5 \lambda$. The left and right columns show propagation for initial fields impinging at the $m=0$ and $m=1$ waveguides, in that order. Solid black lines show the re-normalized intensity at the $n=0$ waveguide and blue dashed lines at the $n=1$ waveguide.}\label{fig:Fig4}
\end{figure}

In order to visualize propagation in the linear $\mathcal{PT}$-symmetric dimer, we will define a re-normalized light intensity at the $n$-th waveguide given an initial field impinging just the $m$-th waveguide in terms of the impulse function,
\begin{eqnarray}
\mathcal{P}_{m,n}(z) = \frac{\vert \mathcal{I}_{m,n}(z) \vert^{2}}{\sum_{j} \vert \mathcal{I}_{m,j}(z) \vert^{2}}.
\end{eqnarray}
The expression $\sum_{j} \vert \mathcal{I}_{m,j}(z) \vert^{2}$ is the total light intensity at $z$ and varies with the propagation distance due to the non-unitary propagation.
Figure \ref{fig:Fig4} shows the re-normalized intensity, the left and right columns show the propagation of a light field impinging at the zeroth and first waveguides, in that order.
The first row, Fig. \ref{fig:Fig4}(a) and \ref{fig:Fig4}(b), shows a device in the $\mathcal{PT}$-symmetric regime, $\gamma < 2 \lambda$, where each field amplitude can be taken as the optical simulation of an harmonic oscillator with periodicity $2 \pi / \Omega_{0}$; note that the re-normalized intensity is periodic but not harmonic.
The second row, Fig. \ref{fig:Fig4}(c) and \ref{fig:Fig4}(d), corresponds to the fully degenerate case, $\gamma = 2 \lambda$, where the field amplitudes simulate free-particle propagation. 
In this case, the device asymptotically balances the re-normalized intensity at each waveguide for any given gain and coupling parameters.
Finally, the third row, Fig. \ref{fig:Fig4}(e) and \ref{fig:Fig4}(f), shows the case where the field amplitudes answer to an inverted oscillator dynamics where the scaled intensity have a more complex asymptotic behavior,
\begin{eqnarray}
\lim_{z \rightarrow \infty}\mathcal{P}_{0,0}(z) &=& \lim_{z \rightarrow \infty}\mathcal{P}_{1,0}(z), \\
&=& \frac{\gamma + \vert \Omega_{0} \vert }{2 \gamma}, \\
\lim_{z \rightarrow \infty}\mathcal{P}_{0,1}(z) &=&  \lim_{z \rightarrow \infty}\mathcal{P}_{1,1}(z), \\
&=& \frac{\gamma - \vert \Omega_{0} \vert }{2 \gamma}.
\end{eqnarray}
These results help us recover the completely degenerate case,
\begin{eqnarray}
\lim_{z \rightarrow \infty} \mathcal{P}_{m,n}(z) = \frac{1}{2}, \quad \gamma = 2 \lambda.
\end{eqnarray}

\section{Conclusion}

In summary, we have shown an optical finite-dimensional representation of the Lorentz group in 2+1D. 
This is a non-compact group, thus, the representation is non-unitary and leads to an array of coupled waveguides with effective linear gain and loses, that is, the imaginary part of the effective refractive index is nonzero.
The resulting photonic lattice is part of the so-called $\mathcal{PT}$-symmetric optical systems and has two distinctive regimes where its dispersion relation is real and pure imaginary.
Nevertheless, it is possible to use a symmetry based approach to construct a closed-form analytic impulse function that can be used to propagate any linear combinations of input fields in any given regime.
We showed that in the regime that keeps the symmetry, with real dispersion relations, the device behaves like an optical oscillator with amplification and attenuation. 
In the broken $\mathcal{PT}$-symmetry regime, the optical simulator delivers a directional coupler with amplification.

As a practical example, we showed that the well known linear $\mathcal{PT}$-symmetric dimer has an underlying $so(2,1)$ symmetry. 
Furthermore, we demonstrated that light propagating through it can act as an optical simulator of an harmonic oscillator, free-particle propagation or inverted oscillator in the $\mathcal{PT}$-symmetric, fully degenerate and broken symmetry regimes, in that order.
Furthermore, the re-normalized intensity has a well behaved asymptotic behavior in the fully degenerate and broken symmetry regimes.



\end{document}